\documentclass[a4paper,11pt]{article}

\pdfoutput=1

\usepackage{hyperref}
\usepackage{lmodern}
\usepackage[utf8]{inputenc}

\usepackage{geometry}

\usepackage{graphicx}
\usepackage{amssymb,amsmath}
\usepackage{multirow}
\usepackage{xcolor}
\usepackage{stmaryrd}

\usepackage{natbib}
\newcommand\pl[1]{\texttt{#1}}
\newcommand\outV{\mathbf{G}}
\newcommand\SBM{Stochastic Block Model}
\newcommand\LBM{Latent Block Model}

\newcommand\intE[2]{{\llbracket#1,#2\rrbracket}}

\newcommand\bR{\mathbf{R}}
\newcommand\bN{\mathbf{N}}
\newcommand\bX{\boldsymbol{X}}
\newcommand\bY{\boldsymbol{Y}}
\newcommand\bZ{\boldsymbol{Z}}
\newcommand\btau{\boldsymbol{\tau}}
\newcommand\balpha{\boldsymbol{\alpha}}
\newcommand\nodeset{\intE{1}{n}}
\newcommand\classset{\intE{1}{Q}}

\newcommand\cF{\mathcal{F}}

\newcommand\ed[2]{{#1}^{(#2)}}
\newcommand\ZA{\ed{Z}{1}}
\newcommand\ZB{\ed{Z}{2}}
\newcommand\tauA{\ed{\tau}{1}}
\newcommand\tauB{\ed{\tau}{2}}
\newcommand\bZA{\ed{\bZ}{1}}
\newcommand\bZB{\ed{\bZ}{2}}
\newcommand\btauA{\ed{\btau}{1}}
\newcommand\btauB{\ed{\btau}{2}}
\newcommand\nA{\ed{n}{1}}
\newcommand\nB{\ed{n}{2}}
\newcommand\QA{\ed{Q}{1}}
\newcommand\QB{\ed{Q}{2}}
\newcommand\alphaA{\ed{\alpha}{1}}
\newcommand\alphaB{\ed{\alpha}{2}}
\newcommand\balphaA{\ed{\balpha}{1}}
\newcommand\balphaB{\ed{\balpha}{2}}
\newcommand\nodesetA{\intE{1}{\nA}}
\newcommand\classsetA{\intE{1}{\QA}}
\newcommand\nodesetB{\intE{1}{\nB}}
\newcommand\classsetB{\intE{1}{\QB}}

\DeclareMathOperator\logit{logit}
\newcommand\ilogit{{\logit}^{-1}}
\newcommand\bbeta{\boldsymbol{\beta}}
\newcommand\bmu{\boldsymbol{\mu}}
\newcommand\bSigma{\boldsymbol{\Sigma}}

\title{Blockmodels: A R-package for estimating in \LBM{} and \SBM{}, with various
    probability functions, with or without covariates.}
\author{J.-B. Leger\footnote{INRA, UR 1404 MaIAGE, Jouy-en-Josas, France}}
\date{}

\begin{document}

\maketitle

\begin{abstract}
    Analysis of the topology of a graph, regular or bipartite one, can be done
    by clustering for regular ones or co-clustering for bipartite ones.

    The \SBM{} and the \LBM{} are two models, which are very similar for respectively
    regular and bipartite graphs, based on probabilistic models.

    Initially developed for binary graphs, these models have been extended to
    valued networks with optional covariates on the edges.

    This paper present a implementation of a Variational EM algorithm for \SBM{}
    and \LBM{} for some common probability functions, Bernoulli, Gaussian and
    Poisson, without or with covariates, with some standard flavors, like
    multivariate extensions. This implementation allow automatic group number
    exploration and selection via the ICL criterion, and allow analyze networks
    with thousands of nodes in a reasonable amount of time.
\end{abstract}

\section{Introduction}

    Complex networks are being more and more studied in different domains such
    as social sciences and biology. Statistical methodology have been developed
    for analysing complex data such as networks or bipartite networks in a way
    that could reveal underlying data patterns through some form of
    classification.

    The models used in this paper are the \SBM, introduced by \citet{nowicki01} and
    \LBM{} introduced by \citet{govaert03} which are the same model for
    regular and bipartite networks.

    This paper introduce a Gnu R package using Variational-EM algorithm as
    introduced by \citet{mariadassou10} for \SBM. The same method is
    applied for \LBM.

    For group number selection, the ICL \citep{biernacki00} is used, and automatic
    exploration is done to found the optimal group number selection.
    This package use automatic reinitialization with other group number result
    to found coherent results.

    The implementation is done in \pl{C++} for CPU intensive step, using the
    linear algebra library \texttt{armadillo} \citep{sanderson10} . The interfacing
    with GNU R  is done by the package \texttt{RcppArmadillo} \citep{eddelbuettel14},
    which itself is using the package \texttt{Rcpp} \citep{eddelbuettel11} to provide
    a easy way to interface \pl{C++} code in packages in GNU R.

    This package is the successor of previous implementation done by the same
    author in \pl{C++} \citep{leger14}. This implementation is more efficient
    for matrix computation, have more available models,
    and more flexibility. This new implementation also have an R interface, and work both for \SBM{} as well for \LBM.

    All non time consuming operations are done in GNU R, and the user interface
    is integrally usable in \pl{R}. This package use parallelism to run
    in the same time on different initialization \textit{via} the
    \texttt{parallel} package in R.

\section{Model framework}

    \SBM{} and \LBM{} models are described with the following notation.

    \subsection{\SBM}

        \subsubsection{Notations}
            Considering the following notation: $\nodeset$ the set of nodes with
            a graph on $n$ nodes. Let $X_{ij}\in\outV$ the weight of the edge
            $(i,j)\in\nodeset^2$. For example, for a binary graph,
            $\outV=\{0,1\}$, and $\bX$ is named the adjacency matrix. Or, for a
            univariate weighed graph $\outV=\bR$, and $\bX$ is named the
            weighted adjacency matrix.  $\bY_{ij}$ the covariates vector
            associated to the edge $(i,j)\in\nodeset^2$, if there is covariates.

            We consider $Q$ classes on nodes, and let $\bZ$ the membership
            matrix defined as $Z_{iq}=1$ if and only if node $i$ is a member of
            class $q$, for $i\in\nodeset$ and $q\in\classset$. Let $\bZ_i$ the
            $i$-th row of the matrix $\bZ$.

            As above, and in all this article, for \SBM, $i$ and $j$ are
            denoting nodes indices (in $\nodeset$), $q$ and $l$ are denoting
            class indices (in $\classset$).

        \subsubsection{Model}
            \paragraph{Latent layer :}
                The class memberships of nodes are driven by independent
                identically distributed multinomial distribution.

                \[
                    \bZ_i \overset{\textrm{i.i.d.}}{\sim} \mathcal{M}(1,
                    \balpha),
                \]
                with $i\in\nodeset$

                Where $\balpha\in\bR_+$ is a parameter as $\sum_{q=1}^Q = 1$.

            \paragraph{Observed layer for \SBM{}:}
                The model is defined by giving the distribution of each edge
                $(i,j)$ conditionally to the membership of node $i$ in the
                $q$-th class and node $j$ in the $l$-th class.

                \[
                    X_{ij}|Z_{iq}Z_{jl}=1 \overset{\textrm{ind}}{\sim}
                    \cF_{ql}^{\bY_{ij}}
                \]
                with $(i,j)\in\nodeset^2$ and $i\neq j$.

                The choice of $\cF$ can lead to a large range of models,
                depending or not on covariates effect.

            \paragraph{Observed layer for symmetric \SBM{}:}
                For symmetric \SBM, edges $(i,j)$ and $(j,i)$ are considered to
                be the same, and only observed one time. The model is the same,
                but only observed of $i<j$:
               
                \[
                    X_{ij}|Z_{iq}Z_{jl}=1 \overset{\textrm{ind}}{\sim}
                    \cF_{ql}^{\bY_{ij}}
                \]
                with $(i,j)\in\nodeset^2$ and $i<j$.

    \subsection{\LBM}
        
        \subsubsection{Notations}
            Considering two types of nodes. The set of the two type are
            $\nodesetA$ and $\nodesetB$, with $\nA$ nodes and $\nB$ nodes of
            each set. The edges only consist of edges between different
            type nodes (\textit{i.e.} for edges
            $(i,j)\in\nodesetA\times\nodesetB$). Let $X_{ij}\in\outV$ be the weight
            of the edges $(i,j)\in\nodesetA\times\nodesetB$. For example, for a
            binary graph, $\outV=\{0,1\}$, and $\bX$ is named the adjacency
            matrix. Or for a univariate weighed graph $\outV=\bR$, and $\bX$ is
            named the weighted adjacency matrix.  $\bY_{ij}$ the covariates vector
            associated to the edge $(i,j)\in\nodesetA\times\nodesetB$, if there
            is covariates.

            We consider $\QA$ classes on nodes for type 1 nodes, and $\QB$
            classes of nodes for type 2 nodes, and let for type 1, $\bZA$ the membership
            matrix defined as $\ZA_{iq}=1$ if and only if node $i$ is a member of
            class $q$, for $i\in\nodesetA$ and $q\in\classsetA$, and $\bZA$ the
            membership matrix defined as $\ZB_{jl}=1$ if and only if node $j$ is
            a member of class $l$, for $j\in\nodesetB$ and $l\in\classsetB$. Let
            $\bZA_i$ the $i$-th row of the matrix $\bZA$, and $\bZB_j$ the
            $j$-th row of the matrix $\bZB$.
           
            As above, and in all this article, for \LBM, $i$ is denoting node
            index of the first type (in $\nodesetA$), $j$ is denoting nodes
            index of the second type (in $\nodesetB$), $q$ is denoting class
            index of the first class ($\classsetA$) and $l$ is denoting
            class index of second type (in $\classsetB$).

        \subsubsection{Model}
            \paragraph{Latent layer :}
            For each type, the node membership are driven by independent
            identically distributed multinomial distribution:

            \[
                \left\{\begin{array}{l}
                        \bZA_i \overset{\textrm{i.i.d.}}{\sim}
                        \mathcal{M}(1,\balphaA), \\
                        \bZB_j \overset{\textrm{i.i.d.}}{\sim}
                        \mathcal{M}(1,\balphaB),
                    \end{array}\right.
            \]
            with $i\in\nodesetA$, $j\in\nodesetB$.

            \paragraph{Observed layer}
                The model is defined by giving the distribution of each edge
                $(i,j)\in\nodesetA\times\nodesetB$ conditionally to the membership of node $i$ in the
                $q$-th class of type 1($q\in\classsetA$) and node $j$ in the
                $l$-th class of type 2 ($l\in\classsetB$).

                \[
                    X_{ij}|\ZA_{iq}\ZB_{jl}=1 \overset{\textrm{ind}}{\sim}
                    \cF_{ql}^{\bY_{ij}}
                \]
                with $(i,j)\in\nodesetA\times\nodesetB$.

                The choice of $\cF$ can lead to a large range of models,
                depending or not on covariates effect.

\section{Estimation procedure}

The used estimation procedure is from \citet{mariadassou10}, with a variational expectation maximization. As
\citet{mariadassou10} the ICL criterion is used for group number selection.

    \subsection{Variational-EM algorithm}
    As done by \citet{mariadassou10}, the
    following criterion is used from a variational approximation of the
    likelihood, for \SBM:

    \[
        J = \sum_{i,q} \tau_{iq}\log(\alpha_q) + \sum_{i,j; i\neq j}\sum_{q,l}
        \tau_{iq}\tau_{jl}\log f_{ql}^{\bY_{ij}}(X_{ij})
    \]

    Where $\btau_i$ is the variational parameter of the multinomial
    distribution which approximate $(\bZ_i|X)$.

    For \LBM, the following criterion is used:

    \[
        J = \sum_{i,q} \tauA_{iq}\log(\alphaA_q) + \sum_{j,l}
        \tauB_{jl}\log(\alphaB_l) + \sum_{i,j}\sum_{q,l}\tauA_{ij}\tauB_{q,l}\log
        f_{ql}^{\bY_{ij}}(X_{ij})
    \]

    The EM with variational approximation is translated is two steps, which are
    repeated until convergence:

    \begin{enumerate}
        \item \textit{Pseudo-E step}: Maximisation with respect to variational
            parameters, $\btau$ for \SBM, and $(\btauA,\btauB)$ for \LBM.
        \item \textit{M-step}: Maximisation with respect to original parameters,
            $\balpha$ and model function parameters for \SBM{} and
            $(\balphaA,\balphaB)$ and model function parameters for \LBM.
    \end{enumerate}

    The maximisation with respect to variational parameters is done by 
    interating a fixed point equation. The maximization with respect to original
    parameters is explicit for $\balpha$ or $(\balphaA,\balphaB)$. For model
    function, the maximization can be done with explicit formula or by a
    numerical maximization algorithm.

    \subsection{ICL criterion}

    For group number selection, the ICL from
    \citet{biernacki00} is used.

    \subsection{Initialization and reinitialization}

    As many algorithm based on the EM algorithm, this method have a huge
    dependency of the initialization quality. Two type of initialization are
    used by this package.

        \paragraph{Absolute Eigenvalues Spectral Clustering :} This variant of
        Spectral Clustering seems to give very good first approximation of a
        classification for \SBM{} case. Furthermore, of \SBM, with Bernoulli
        distribution without covariate, the Absolute Eigenvalues Spectral
        Clustering is consistent, see \citet{rohe11}.
        For small graphs (less than 1000 nodes), the obtained clustering seems
        not to be a very good results to be used as is, but it is a good start
        point for this package.

        To take care of covariates, where there are ones, the Absolute
        Eigenvalues Spectral clustering is run on the residual graphs of the
        regression (which is in fact the residual graphs for the one-group
        model).

        For \SBM, this residual graphs is directly used as a input of the
        Absolute Eigenvalues Spectral Clustering. For \LBM, the residual graph
        is projected on each node type, and a Absolute Eigenvalues Spectral
        Clustering is done for each node type.

        \paragraph{Reinitialization :} The obtained results by the method for a
        group number is used to provide new initialization for previous groups
        number (by merging groups) and next groups number (by splitting groups).

        Due to the high number of reinitialization proposed,
        in some case, the criterion is evaluated on each provided
        reinitializations, and only best ones are used (the number of used iterations
        each step is depending of a constant and the group number, the constant
        can be changed by user).

        The process of reinitialization is done while reinitialization improve
        the criterion.

    \subsection{Group number exploration}

        To explore the group number, the model is run for a beginning set of
        group number (which can be changed by the user). After that, for the
        selected number of group (the ICL maximum), the exploration is done to a
        maximal number of groups which is a constant (by default $1.5$, user
        modifiable) times the selected number of groups.

        It is important to explore after the maximum, oversplitted groups (after
        the ICL maximum) can provide good reinitialization by merging, and
        change the maximum location.

\section{Architecture}

    In this section the architecture of the package is described. The package
    use \pl{C++} for CPU intensive operations and \pl{R} for other operations.
    The general architecture is describe in the following section and each type
    of code is describe below.

    \subsection{General architecture}

    The package is usable inside GNU R, therefore the interface of the package
    is in \pl{R}, user provide data to the package in R, and the time-consuming
    operation is written in \pl{C++}.

    \paragraph{Model definition :} The user define a object with the model, \SBM{}
    or \LBM{} and model function, and the data, adjacency of the network and
    covariates if there are ones.

    The returned object have methods which provide estimation and access to the
    results.

    \subsection{\pl{R} code}
        
        All the \pl{R} code use \texttt{RefClass} (S4) class. These type of class,
        are the equivalent in \pl{R} of class in most other programming language
        where the methods of the class can modify the object itself.

        All estimation for different initialization of a number of group are run
        in parallel, via the \texttt{parallel} GNU R package on platforms which
        support parallelism (Linux, Solaris, *BSD, MacOS). On Windows,
        estimations for different initialization are run sequentially because
        the parallel package does not support this OS.

        \subsubsection{Memberships}

        The memberships, \SBM{} or \LBM, have specific functions, for estimation
        and for results access. They inherit from a virtual class membership
        which if the one considered by other functions.

        \subsubsection{Model functions}

        All model functions inherit from a virtual class which is used by other
        functions. Some model functions inherit between themselves when a model
        function is the extension of another. This code contains non-time
        consuming model specific functions, as normalization, or displaying
        functions.

        A example file is given in the source code which indication how to write
        a model specific class.

    \subsection{\pl{C++} code}

        The EM is implemented in C++, via templated function. Generic
        template functions are written, and the EM function is evaluated at
        compilation for each model function and each membership type (\LBM{} or
        \SBM).

        Templated function are defined, with generic code. Each model need
        template specialization to specify the model functions. The
        specialization can be done in higher lever when vectorized function
        exists (for the fixed point equation, for explicit maximum in the M step
        if there is one, or for the gradient calculation) or can be done in the
        lower level with only providing the model functions and derivatives.

        An example file is given in the source code which indicating the function
        to specialize to describe a new model function.

\section{Implemented model functions}

    Common model functions are implemented, they are described below. Some model
    have vectorized specialization to provide fast code when this is possible.

    \subsection{Bernoulli family}
        \subsubsection{Bernoulli}

        This is the common \SBM{} or \LBM. Links are valued in $\outV=\{0,1\}$.
        The model is defined as below:

        \[
            \cF_{ql}^{\bY_{ij}} = \mathcal{B}(\pi_{ql})
        \]

        Parameters:
        \begin{itemize}
            \item $\pi_{ql}\in[0,1]$, $(q,l)\in\classset^2$ for \SBM{} or
                $(q,l)\in\classsetA\times\classsetB$ for \LBM. This is the group
                effect.
        \end{itemize}

        The implementation is vectorized for the E-step, and have an explicit
        maximum computed with vectorized formula in the M-step.

        The model is accessible by \texttt{BM\_bernoulli}

        \subsubsection{Bernoulli multiplex}

        This model is a multivariate non-independent Bernoulli distribution.
        Links are valued in $\outV=\{0,1\}^p$.

        The model is defined as follow for \SBM:
        
        \[
            \forall x\in\{0,1\}^p\qquad
            \mathbf{P}\left(X_{ij}=x|Z_{iq}Z_{jl}=1\right) = \pi_{ql}[x]
        \]
        
        and for \LBM:
        
        \[
            \forall x\in\{0,1\}^p\qquad
            \mathbf{P}\left(X_{ij}=x|\ZA_{iq}\ZB_{jl}=1\right) = \pi_{ql}[x]
        \]

        Parameters:
        \begin{itemize}
            \item $\pi_{ql}[x]\in[0,1]$ $(q,l)\in\classset^2$ for \SBM{} or
                $(q,l)\in\classsetA\times\classsetB$ for \LBM, $x\in\{0,1\}^p$,
                under the constraint $\forall q,l; \sum_x \pi_{ql}[x]=1$. This
                is the group effect.
        \end{itemize}

        The implementation is vectorized for the E-step, and have a explicit
        maximum computed with vectorized formula in the M-step.

        The model is accessible by \texttt{BM\_bernoulli\_multiplex}

        \subsubsection{Bernoulli with covariates}

        This model provide a logistic regression with a group effect  which is
        the intercept and a covariates effect.

        Links are valued in $\outV=\{0,1\}$. The model is defined as below:

        \[
            \cF_{ql}^{\bY_{ij}} = \mathcal{B}\left(\ilogit(m_{ql}+\bbeta^T\bY_{ij})\right)
        \]

        where $\logit(p)=\log\left(\frac{p}{1-p}\right)$.

        Parameters:
        \begin{itemize}
            \item $m_{ql}\in\bR$, $(q,l)\in\classset^2$ for \SBM{} or
                $(q,l)\in\classsetA\times\classsetB$ for \LBM. This is the group
                effect.
            \item $\beta$, the covariates effect.
        \end{itemize}

        Two implementation of this model is describe below.

            \paragraph{Standard implementation :}

            Due to the non-separability between the
            group effect and the covariates effect, even in polynomial form, the specialization must be
            done in the lower level. This implementation is very slow.

            The E-step is not vectorized, the maximum is numerically computed in
            the M-step without vectorized gradient calculation.

            This implementation is accessible by
            \texttt{BM\_bernoulli\_covariates}.

            \paragraph{Fast implementation with approximation :}

            Alternatively to the previous implementation, which is exact, a
            fast implementation is provided using a approximation.

            Let $g$ defined as:
            \[
                g: x\mapsto \frac{1}{2}x +
                \log\left(1-\frac{1}{1+\exp(-x)}\right)
            \]

            With this function, the log likelihood for a link can be expressed
            as:

            \[
                \log f_{ql}^{\bY_{ij}}(X_{ij}) = 
                \left(X_{ij}-\frac{1}{2}\right)\left(m_{ql}+\bbeta^T\bY_{ij}\right)
                + g\left(m_{ql}+\bbeta^T\bY_{ij}\right)
            \]

            Terms using $\log f_{ql}^{\bY_{ij}}(X_{ij})$ and variational
            parameters are summed over $i,j$ and $q,l$. As for above, due to the
            form of $g$, this sum can not be separated.
            
            To separate the sum and
            vectorize the computation, $g$ is substituted by a polynomial which
            approximate the function. The polynomial involve powers of the term
            $m_{ql}+\bbeta^T\bY_{ij}$. Terms are separated by power of $m_{ql}$
            and $\bbeta^T\bY_{ij}$ using the Binomial theorem.

            By changing the summing order, and
            considering the sum over terms as lower priority, we can separate
            and vectorize the computation for each term, involving only powers
            of $m_{ql}$ and $\bbeta^T\bY_{ij}$.

            The function $g$ is even, so the polynomial approximation involve
            only even power terms. The polynomial is chosen of degree $14$, in order to
            approximate the best the function $g$ on $[-15,15]$. The function
            $g$ is concave, though the polynomial approximating of the function $g$ does not 
            need to be concave to be a good approximation. Still an upper bound
            constraint is added on the second derivative. This last constraint
            provide a good numerical stability with a very small approximation
            loss.

            In simulation, for all $i,j,q,l$, if $m_{ql}+\bbeta^T\bY_{ij} \in
            [-15,15]$, the approximation method give the same results as the
            exact method.

            Due to asymptotic branch of the polynomial which go very quickly to
            $-\infty$, in general case, this fast method have the same behavior
            of the logistic regression under the constraint:

            \[
                \forall i,j,q,l \left|m_{ql}+\bbeta^T\bY_{ij}\right|\leq15
            \]

            The E-step is vectorized, the maximum is numerically computed in the
            M-step with a vectorized gradient calculation.

            This implementation is accessible by
            \texttt{BM\_bernoulli\_covariates\_fast}.

    \subsection{Gaussian family}
        \subsubsection{Gaussian}

        This is the \SBM{} or \LBM{} with normally distributed values on links.
        Links are valued in $\outV=\bR$.

        The model is defined as below:

        \[
            \cF_{ql}^{\bY_{ij}} = \mathcal{N}(\mu_{ql},\sigma^2)
        \]

        Parameters:
        \begin{itemize}
            \item $\mu_{ql}\in\bR$, $(q,l)\in\classset^2$ for \SBM{} or
                $(q,l)\in\classsetA\times\classsetB$ for \LBM. This is the group
                effect.
            \item $\sigma^2$, the parameter of the variance.
        \end{itemize}

        The implementation is vectorized for the E-step, and have a explicit
        maximum computed with vectorized formula in the M-step.

        The model is accessible by \texttt{BM\_gaussian}

        \subsubsection{Gaussian multivariate}

        This is the \SBM{} or \LBM{} with multivariate normally distributed values
        on links. Links are valued in $\outV=\bR^p$.

        The model is defined as below:

        \[
            \cF_{ql}^{\bY_{ij}} = \mathcal{N}(\bmu_{ql},\bSigma)
        \]

        Parameters:
        \begin{itemize}
            \item $\bmu_{ql}\in\bR^p$, $(q,l)\in\classset^2$ for \SBM{} or
                $(q,l)\in\classsetA\times\classsetB$ for \LBM. This vector is the group
                effect.
            \item $\bSigma$, the variance-covariance matrix.
        \end{itemize}

        Three flavors of this model are provided depending of the shape of the
        variance covariance matrix.

        All the flavors have vectorized E-step and a explicit maximum with
        vectorized computation in the M-step.

            \paragraph{Independent homoscedastic case}:
            This case considers the components are independent and have same
            variance. We consider $\bSigma = \sigma^2 I_p$ where $I_p$ is the
            identity matrix of size $p$.

            This model is accessible by
            \texttt{BM\_gaussian\_multivariate\_independent\_homoscedastic}.
            
            \paragraph{Independent case}:
            
            This case considers the components are independent and have same
            variance. We consider $\bSigma$ is a diagonal matrix.

            This model is accessible by
            \texttt{BM\_gaussian\_multivariate\_independent}.

            \paragraph{General case}
            
            We only assume that $\bSigma$ is
            semi-definite positive matrix, which is contained in the likelihood.

            This model is accessible by
            \texttt{BM\_gaussian\_multivariate}.

        \subsubsection{Gaussian with covariates}

        This model is a standard linear regression on the covariates for \SBM{}
        and \LBM. Links are valued on $\outV=R$. The model is defined as below:

        \[
            \cF_{ql}^{\bY_{ij}} = \mathcal{N}\left( \mu_{ql} + \bbeta^T\bY_{ij},
            \sigma^2 \right)
        \]

        Parameters:
        \begin{itemize}
            \item $\mu_{ql}\in\bR$, $(q,l)\in\classset^2$ for \SBM{} or
                $(q,l)\in\classsetA\times\classsetB$ for \LBM. This is the group
                effect.
            \item $\bbeta$, the covariates effect,
            \item $\sigma^2$ the parameter of variance.
        \end{itemize}

        This model have vectorized E-step, a numerically maximum with vectorized
        computation in the M-step.

        This model is accessible by \texttt{BM\_gaussian\_covariates}

    \subsection{Poisson family}
        \subsubsection{Poisson}

        This is the \SBM{} or \LBM{} with Poisson distributed values on links.
        Links are valued in $\outV=\bN$.

        The model is defined as below:

        \[
            \cF_{ql}^{\bY_{ij}} = \mathcal{P}(\lambda_{ql})
        \]

        Parameters:
        \begin{itemize}
            \item $\lambda_{ql}\in\bR$, $(q,l)\in\classset^2$ for \SBM{} or
                $(q,l)\in\classsetA\times\classsetB$ for \LBM. This is the group
                effect.
        \end{itemize}

        The implementation is vectorized for the E-step, and have a explicit
        maximum computed with vectorized formula in the M-step.

        The model is accessible by \texttt{BM\_poisson}

        \subsubsection{Poisson with covariates}

        This model is a Poisson regression on the covariates for \SBM{} and \LBM.
        Links are valued on $\outV=\bN$. The model is defined as below:

        \[
            \cF_{ql}^{\bY_{ij}} =
            \mathcal{P}\left(\lambda_{ql}\exp(\bbeta^T\bY_{ij})\right)
        \]
        
        Parameters:
        \begin{itemize}
            \item $\lambda_{ql}\in\bR$, $(q,l)\in\classset^2$ for \SBM{} or
                $(q,l)\in\classsetA\times\classsetB$ for \LBM. This is the group
                effect.
            \item $\bbeta$, the covariates effect,
        \end{itemize}

        This model have vectorized E-step, a numerically maximum with vectorized
        computation in the M-step.

        This model is accessible by \texttt{BM\_poisson\_covariates}

\section{Execution time on a example}
    \subsection{Methodology}

    All this tests are done with \SBM.

    For each model, a network is simulated accordingly, in the way
    documented in manuals.

    Four conditions are simulated: 
    \begin{itemize}
        \item 5 groups and 100 nodes
        \item 5 groups and 200 nodes
        \item 10 groups and 100 nodes
        \item 10 groups and 200 nodes.
    \end{itemize}
   
    In order for the result to be comparable, 
    the automatic group number exploration is disabled, and the exploration is
    force to explore all groups number between 1 and twice the number of
    simulated groups.

    Each estimation is repeated 5 times, the median is the result retained and reported in the table below. The reported time is the CPU time which
    cumulates the execution time of all parallel process. The real execution time
    is less than the CPU time, due to parallelism. All computation are run on
    the same machine, with a Intel Xeon X5675 CPU.

    \subsection{Results}
    
    The results are:

    {\scriptsize

    \begin{tabular}{|c|c|r|r|r|r|}
        \cline{3-6}
        \multicolumn{2}{c|}{} & \multicolumn{2}{|c|}{100 nodes} & \multicolumn{2}{|c|}{200 nodes} \\
        \cline{3-6}
        \multicolumn{2}{c|}{} & 5 groups & 10 groups & 5 groups & 10 groups \\
        \hline
        \multirow{4}{*}{Bernoulli}&Standard                      &          10 s &            57 s &           16 s &        3 m 30 s \\
                                  &Multiplex                     &           9 s &        1 m 37 s &           43 s &        3 m 57 s \\
                                  &Covariates (exact)            & 4 h 56 m 49 s & 135 h 55 m 03 s & 11 h 38 m 35 s & 761 h 53 m 40 s \\
                                  &Covariates (fast)             &      5 m 29 s &   3 h 10 m 58 s &      38 m 00 s &  16 h 01 m 32 s \\
        \hline
        \multirow{5}{*}{Gaussian} &Standard                      &           8 s &        1 m 07 s &           19 s &        2 m 51 s \\
                                  &Multivariate (indep. homosc.) &          11 s &        1 m 40 s &           28 s &        2 m 49 s \\
                                  &Multivariate (indep.)         &          10 s &            58 s &           37 s &        2 m 52 s \\
                                  &Multivariate                  &           5 s &            57 s &           10 s &        1 m 03 s \\
                                  &Covariates                    &          51 s &   1 h 27 m 46 s &       2 m 59 s &   1 h 40 m 25 s \\
        \hline
        \multirow{2}{*}{Poisson}  &Standard                      &          10 s &        1 m 25 s &           27 s &        4 m 23 s \\
                                  &Covariates                    &      1 m 17 s &   2 h 41 m 16 s &       4 m 06 s &   3 h 49 m 14 s \\
        \hline
    \end{tabular}
}
    
    The results are highly dependent on the network and the signal to noise ratio.
    Therefore, no comparison should be done between families, which use
    different models to simulate networks.

    Users should keep in mind that above times have been computed on
    simulated graphs generated from the \textit{true} model. Thus these
    timings are provided only as a guide. Times for applications on real
    graph may differ.

\section{Application on the Debian keyring signing network}
    \subsection{The Debian project}
    \subsection{The data}
    \subsection{Estimation procedure}
    \subsection{Results}

\section*{References}
\bibliography{biblio}

\end{document}